\documentclass[12pt,a4paper]{article}
\usepackage{amsmath,amsfonts,amsthm,graphicx,lscape}
\usepackage[dvips]{psfrag}
\usepackage{cite}

\usepackage[normalem]{ulem}

\usepackage{textcomp}
\usepackage{amsmath,amsthm,amssymb,mathrsfs,cases}
\usepackage{amsfonts,graphicx,lscape}

\newcommand{\vt}[1]{\mbox{\boldmath$#1$}}
\newcommand{\sca}[2]{\langle #1, #2 \rangle}

\usepackage{accents}
\makeatletter
\def\widebar{\accentset{{\cc@style\underline{\mskip10mu}}}}
\makeatother

\numberwithin{equation}{section} 
\newtheorem{theorem}{Theorem}[section]
\newtheorem{proposition}[theorem]{Proposition}

\def\beqa{\begin{eqnarray}}
\def\enqa{\end{eqnarray}}
\def\beq{\begin{equation}}
\def\enq{\end{equation}}

\begin{document}
\title{
Integrability of
a discrete Yajima--Oikawa system
}
\author{Takayuki \textsc{Tsuchida}
}
\maketitle
\begin{abstract} 
A space discretization of an integrable 
long wave--short wave interaction model, 
called the Yajima--Oikawa system, was 
proposed 
in the recent paper arXiv:1509.06996 
using the Hirota bilinear method
(see also https://link.aps.org/doi/10.1103/PhysRevE.91.062902). 
In this paper, we propose a Lax-pair representation 
for the discrete Yajima--Oikawa system
as well as its multicomponent generalization 
also considered in arXiv:1509.06996 
and 
prove that it has an infinite number of 
conservation laws. 
We 
also 
derive
the 
next higher 
flow of 
the 
discrete Yajima--Oikawa hierarchy, 
which 
generalizes 
a modified version 
of the 
Volterra lattice. 
Relations to two integrable 
discrete nonlinear Schr\"odinger hierarchies, 
the Ablowitz--Ladik hierarchy
and the Konopelchenko--Chudnovsky hierarchy, 
are clarified. 
\end{abstract}

%
%

\noindent
\tableofcontents

\newpage
\section{Introduction}

The 
system of two coupled partial differential equations 
\begin{subnumcases}{\label{YO}}
\mathrm{i} S_t = S_{xx} 
+  
LS, 
\label{YO-1} \\[1pt]
L_t = 2 \left( \left| S \right|^2 \right)_x, 
\label{YO-2} 
\end{subnumcases}
where the subscripts denote 
the partial differentiation, 
describes the interaction 
between the complex-valued short-wave component 
$S$
and the real-valued long-wave component 
$L$ 
and is completely integrable. 
We call (\ref{YO})
the Yajima--Oikawa system, 
because 
a system equivalent to (\ref{YO}), up to a 
simple transformation, 
was first solved by Yajima and Oikawa using the inverse scattering method~\cite{YO76}. 
In analogy with the 
multicomponent generalization of the 
cubic nonlinear Schr\"odinger equation~\cite{Mana74,YO2,Ab78,New79,Mak82}, 
a multicomponent generalization of the Yajima--Oikawa system~\cite{YCMa81, Mel83,MPK83},
\begin{subnumcases}{\label{vYO}}
\mathrm{i} \vt{S}_t = \vt{S}_{xx} 
+ L \vt{S}, 
\label{vYO-1} \\[1pt]
L_t = 2 \sca{\vt{S}}{\Sigma \vt{S}^\ast}_x, 
\label{vYO-2} 
\end{subnumcases}
can be 
considered. 
Here, $\vt{S}$ is 
a column 
vector 
composed of short-wave components,  \mbox{$\sca{\,\cdot\,}{\,\cdot\,}$} 
represents the standard 
scalar product, 
the asterisk denotes the complex conjugation 
and $\Sigma$ is a diagonal matrix with each diagonal entry $+1$ or $-1$. 
In this paper, 
the product of 
a scalar $s$ and a column 
vector $\vt{V}$ 
is expressed as either $s \vt{V}$ or $\vt{V} s$, i.e., 
the order in the product 
is irrelevant. 
Note that any nonsingular Hermitian matrix $\Sigma$ can be 
reduced to this diagonal form 
using 
a linear transformation acting on the vector variable $\vt{S}$. 
The multicomponent Yajima--Oikawa system (\ref{vYO}) 
is 
also 
completely 
integrable 
and 
possesses 
exact multisoliton solutions~\cite{YCMa81, Mel83, MPK83,Dub88, Cheng}.  

In the recent paper~\cite{Maruno16}, a space discretization of 
the multicomponent Yajima--Oikawa system (\ref{vYO}), 
\begin{subnumcases}{\label{dvYO}}
\mathrm{i} \vt{S}_{n,t} = \left( 1+\frac{1}{2} L_n \right)  \left( \vt{S}_{n+1}+\vt{S}_{n-1} \right), 
\label{dvYO-1} \\[1pt]
\left[ \log \left( 1+ \frac{1}{2} L_n \right) \right]_t = \frac{1}{2} \hspace{1pt} \boldsymbol{\Delta}_n^+ 
\left( \sca{\vt{S}_{n}}{\Sigma\vt{S}_{n-1}^\ast} + \sca{\vt{S}_{n-1}}{\Sigma\vt{S}_{n}^\ast} \right), 
\label{dvYO-2} 
\end{subnumcases}
where 
$\boldsymbol{\Delta}_n^+$ 
is the forward difference operator, i.e.,  
\mbox{$ \boldsymbol{\Delta}_n^+ f_{n} 
:=
f_{n+1}
-f_{n}$}, 
was proposed using the Hirota bilinear method and 
multisoliton solutions in terms of pfaffians 
were constructed; 
we should also mention that 
a \mbox{$(2+1)$}-dimensional 
generalization~\cite{Mel83,Zakh,Nizh82} 
of the discrete Yajima--Oikawa system
was studied previously in~\cite{Yu15}.  
We set 
\mbox{$
1+ \frac{1}{2} L_n 
=:v_n 
$}
and consider a slightly generalized version of (\ref{dvYO}): 
\begin{subnumcases}{\label{dgvYO}}
\mathrm{i} \vt{S}_{n,t} = v_n \left( \vt{S}_{n+1}+\vt{S}_{n-1} \right) -c \vt{S}_{n}, 
\label{dgvYO-1} \\[1pt]
\mathrm{i} \widebar{\vt{S}}_{n,t} = -v_n \left( \widebar{\vt{S}}_{n+1}+\widebar{\vt{S}}_{n-1} \right)
	+ c \widebar{\vt{S}}_{n}, 
\label{dgvYO-2} \\[1pt]
v_{n,t} = \frac{1}{2} v_n \hspace{1pt} \boldsymbol{\Delta}_n^+ 
\left( \sca{\vt{S}_{n}}{\widebar{\vt{S}}_{n-1}} + \sca{\vt{S}_{n-1}}{\widebar{\vt{S}}_{n}} \right), 
\label{dgvYO-3} 
\end{subnumcases}
where $c$ is an arbitrary constant and $\widebar{\vt{S}}_{n}$ 
may or 
may not be 
related to 
the complex conjugate of $\vt{S}_{n}$. 

The main 
objective 
of this paper 
is to demonstrate 
the 
integrability of 
the 
discrete multicomponent Yajima--Oikawa system (\ref{dgvYO}). 
In section 2, 
we 
prove that  (\ref{dgvYO}) admits 
a Lax-pair representation~\cite{Lax} and 
construct the next higher flow of the discrete multicomponent Yajima--Oikawa 
hierarchy 
that can be reduced to 
a modified version of the Volterra lattice. 
We also 
discuss 
the continuous limit of space. 
In section 3, we show that  (\ref{dgvYO}) 
possesses an infinite set of conservation laws. 
In section 4, we 
show how  (\ref{dgvYO}) can be 
reduced to elementary flows of 
two integrable 
discrete nonlinear Schr\"odinger hierarchies:\ 
the Ablowitz--Ladik hierarchy~\cite{AL1} and 
the Konopelchenko--Chudnovsky hierarchy~\cite{Kono82,Chud1,Chud2}. 
Section 5 is devoted to 
concluding remarks. 

\section{Lax pair}

The main result of this paper is the following proposition, 
which provides 
an auxiliary linear problem for (\ref{dgvYO}), 
generally called the Lax pair~\cite{Lax}. 
\begin{proposition}
\label{prop2.1}
The discrete multicomponent Yajima--Oikawa system \mbox{$(\ref{dgvYO})$} 
is equivalent to the 
compatibility condition 
of 
the overdetermined linear equations 
for  $\psi_n$, $\vt{\phi}_n$ and $\vt{\chi}_n$:
\begin{subnumcases}{\label{YO-L}}
v_{n} \left( \psi_{n+1} + \psi_{n-1} \right) = \lambda \psi_{n} 
	-\sca{\vt{S}_n}{\vt{\phi}_n}-\sca{\vt{\chi}_n}{\widebar{\vt{S}}_{n}}, 
\label{YO-L1}
\\[1pt]
\vt{\phi}_{n+1} -\vt{\phi}_n = \frac{\mathrm{i}}{2} \widebar{\vt{S}}_{n} \left( \psi_{n+1} + \psi_{n-1} \right),
\label{YO-L2}
\\[1pt]
\vt{\chi}_{n+1} +\vt{\chi}_n = \frac{\mathrm{i}}{2}  \vt{S}_{n}  \left( \psi_{n+1} + \psi_{n-1} \right),
\label{YO-L3}
\end{subnumcases}
and 
\begin{subnumcases}{\label{YO-M}}
\mathrm{i} \psi_{n,t} = v_{n} \left( \psi_{n+1} + \psi_{n-1} \right) - c \psi_n, 
\label{YO-M1}
\\[1pt]
\vt{\phi}_{n,t} = \frac{1}{2} v_n \widebar{\vt{S}}_{n-1} \left( \psi_{n+1} + \psi_{n-1} \right) 
 - \frac{1}{2} v_{n-1} \widebar{\vt{S}}_{n} \left( \psi_{n} + \psi_{n-2} \right), 
\\[1pt]
\vt{\chi}_{n,t} = \frac{1}{2} v_n \vt{S}_{n-1} \left( \psi_{n+1} + \psi_{n-1} \right) 
 + \frac{1}{2} v_{n-1} \vt{S}_{n} \left( \psi_{n} + \psi_{n-2} \right) + 2 \mathrm{i} c \vt{\chi}_n, \;\;\;\;\;
\label{YO-M3}
\end{subnumcases}
where $\lambda$ is 
a constant spectral parameter. 
\end{proposition}

Proposition~\ref{prop2.1} can be proved by a straightforward calculation; 
we 
operate with $\partial_t$ on 
(\ref{YO-L1}),  
(\ref{YO-L2}) and (\ref{YO-L3}) and use (\ref{YO-M}) and (\ref{YO-L1}) 
to remove the time derivative and the spectral parameter. 
The resulting three 
equations 
are 
\[
\begin{cases}
\mathrm{i} \left[ 
v_{n,t} - \frac{1}{2} v_n \hspace{1pt} \boldsymbol{\Delta}_n^+ 
\left( \sca{\vt{S}_{n}}{\widebar{\vt{S}}_{n-1}} + \sca{\vt{S}_{n-1}}{\widebar{\vt{S}}_{n}} \right)
\right] \left( \psi_{n+1} + \psi_{n-1} \right)
\nonumber \\
\quad \mbox{} + \sca{\mathrm{i} \vt{S}_{n,t} - v_n \left( \vt{S}_{n+1}+\vt{S}_{n-1} \right) + c \vt{S}_{n}}{\vt{\phi}_n}
\nonumber \\
\quad \mbox{} + \sca{\vt{\chi}_n}{\mathrm{i} \widebar{\vt{S}}_{n,t} + v_n \left( \widebar{\vt{S}}_{n+1}+\widebar{\vt{S}}_{n-1} \right)
	- c \widebar{\vt{S}}_{n}}
 = 0, 
\nonumber \\[5pt]
\left[ \mathrm{i} \widebar{\vt{S}}_{n,t} + v_n \left( \widebar{\vt{S}}_{n+1}+\widebar{\vt{S}}_{n-1} \right)
	- c \widebar{\vt{S}}_{n} \right] \left( \psi_{n+1} + \psi_{n-1} \right) = \vt{0}, 
\nonumber \\[5pt]
\left[ 
\mathrm{i} \vt{S}_{n,t} - v_n \left( \vt{S}_{n+1}+\vt{S}_{n-1} \right) + c \vt{S}_{n}
\right]  \left( \psi_{n+1} + \psi_{n-1} \right) = \vt{0}, 
\nonumber
\end{cases}
\]
which are 
indeed 
equivalent to 
(\ref{dgvYO}). 

The scalar $\psi_n$ and the column 
vectors $\vt{\phi}_n$ and $\vt{\chi}_n$ 
constitute the linear eigenfunction in this Lax-pair representation. 
Note that 
the time-evolution equation (\ref{YO-M1}) for $\psi_n$ 
takes the same form as the original 
time-evolution equation  (\ref{dgvYO-1}) for $\vt{S}_n$, 
which is a peculiar feature of the Lax-pair representations 
for Yajima--Oikawa-type systems. 
%
%
%

By changing the time evolution of the linear eigenfunction 
appropriately, 
we obtain 
the next higher flow of 
the discrete multicomponent Yajima--Oikawa hierarchy. 
%
\begin{proposition}
\label{prop2.2}
The compatibility condition 
of the overdetermined linear equations 
for $\psi_n$, $\vt{\phi}_n$ and $\vt{\chi}_n$: 
\[
\begin{cases}
v_{n} \left( \psi_{n+1} + \psi_{n-1} \right) = \lambda \psi_{n} 
	-\sca{\vt{S}_n}{\vt{\phi}_n}-\sca{\vt{\chi}_n}{\widebar{\vt{S}}_{n}}, 
\\[1pt]
\vt{\phi}_{n+1} -\vt{\phi}_n = \frac{\mathrm{i}}{2} \widebar{\vt{S}}_{n} \left( \psi_{n+1} + \psi_{n-1} \right),
\\[1pt]
\vt{\chi}_{n+1} +\vt{\chi}_n = \frac{\mathrm{i}}{2}  \vt{S}_{n}  \left( \psi_{n+1} + \psi_{n-1} \right),
\end{cases}
\]
and 
\[
\left\{
\begin{split}
\psi_{n,\tau} &= v_{n} v_{n+1} \left( \psi_{n+2} + \psi_{n} \right) - v_{n} v_{n-1} \left( \psi_{n} + \psi_{n-2} \right) 
\nonumber \\ 
& \hphantom{=} \mbox{} + \frac{\mathrm{i}}{2} v_{n} \left( \sca{\vt{S}_{n+1}}{\widebar{\vt{S}}_{n}} 
+ \sca{\vt{S}_{n}}{\widebar{\vt{S}}_{n+1}} + \sca{\vt{S}_{n}}{\widebar{\vt{S}}_{n-1}} + \sca{\vt{S}_{n-1}}{\widebar{\vt{S}}_{n}} 
\right) \left( \psi_{n+1} + \psi_{n-1} \right), 
\\[2pt]
\vt{\phi}_{n,\tau} &= \frac{\mathrm{i}}{2} v_n v_{n+1} \widebar{\vt{S}}_{n-1} \left( \psi_{n+2} + \psi_{n} \right) 
 + \frac{\mathrm{i}}{2} v_n v_{n-1} \widebar{\vt{S}}_{n-2} \left( \psi_{n+1} + \psi_{n-1} \right)
\nonumber \\ 
& \hphantom{=} \mbox{} + \frac{\mathrm{i}}{2} v_n v_{n-1} \widebar{\vt{S}}_{n+1} \left( \psi_{n} + \psi_{n-2} \right) 
 + \frac{\mathrm{i}}{2} v_{n-1} v_{n-2} \widebar{\vt{S}}_{n} \left( \psi_{n-1} + \psi_{n-3} \right)
\nonumber \\ 
& \hphantom{=} \mbox{} -\frac{1}{4} v_{n} \left( \sca{\vt{S}_{n+1}}{\widebar{\vt{S}}_{n}} 
+ \sca{\vt{S}_{n}}{\widebar{\vt{S}}_{n+1}} + \sca{\vt{S}_{n}}{\widebar{\vt{S}}_{n-1}} + \sca{\vt{S}_{n-1}}{\widebar{\vt{S}}_{n}} 
\right) \widebar{\vt{S}}_{n-1} \left( \psi_{n+1} + \psi_{n-1} \right)
\nonumber \\ 
& \hphantom{=} \mbox{} +\frac{1}{4} v_{n-1} \left( \sca{\vt{S}_{n}}{\widebar{\vt{S}}_{n-1}} 
+ \sca{\vt{S}_{n-1}}{\widebar{\vt{S}}_{n}} + \sca{\vt{S}_{n-1}}{\widebar{\vt{S}}_{n-2}} + \sca{\vt{S}_{n-2}}{\widebar{\vt{S}}_{n-1}} 
\right) \widebar{\vt{S}}_{n} \left( \psi_{n} + \psi_{n-2} \right), 
\\[2pt]
\vt{\chi}_{n,\tau} &= \frac{\mathrm{i}}{2} v_n v_{n+1} \vt{S}_{n-1} \left( \psi_{n+2} + \psi_{n} \right) 
 - \frac{\mathrm{i}}{2} v_n v_{n-1} \vt{S}_{n-2} \left( \psi_{n+1} + \psi_{n-1} \right)
\nonumber \\ 
& \hphantom{=} \mbox{} + \frac{\mathrm{i}}{2} v_n v_{n-1} \vt{S}_{n+1} \left( \psi_{n} + \psi_{n-2} \right) 
 - \frac{\mathrm{i}}{2} v_{n-1} v_{n-2} \vt{S}_{n} \left( \psi_{n-1} + \psi_{n-3} \right)
\nonumber \\ 
& \hphantom{=} \mbox{} -\frac{1}{4} v_{n} \left( \sca{\vt{S}_{n+1}}{\widebar{\vt{S}}_{n}} 
+ \sca{\vt{S}_{n}}{\widebar{\vt{S}}_{n+1}} + \sca{\vt{S}_{n}}{\widebar{\vt{S}}_{n-1}} + \sca{\vt{S}_{n-1}}{\widebar{\vt{S}}_{n}} 
\right) \vt{S}_{n-1} \left( \psi_{n+1} + \psi_{n-1} \right)
\nonumber \\ 
& \hphantom{=} \mbox{} -\frac{1}{4} v_{n-1} \left( \sca{\vt{S}_{n}}{\widebar{\vt{S}}_{n-1}} 
+ \sca{\vt{S}_{n-1}}{\widebar{\vt{S}}_{n}} + \sca{\vt{S}_{n-1}}{\widebar{\vt{S}}_{n-2}} + \sca{\vt{S}_{n-2}}{\widebar{\vt{S}}_{n-1}} 
\right) \vt{S}_{n} \left( \psi_{n} + \psi_{n-2} \right),
\end{split}
\right.
\]
where $\lambda$ is 
a constant spectral parameter, is equivalent to the system:
\begin{subnumcases}{\label{dgvhYO}}
\vt{S}_{n,\tau} = v_n v_{n+1} \left( \vt{S}_{n+2}+\vt{S}_{n} \right) 
	- v_n v_{n-1} \left( \vt{S}_{n}+\vt{S}_{n-2} \right)
\nonumber \\ \hspace{11mm}
\mbox{} + \frac{\mathrm{i}}{2} v_{n} \left( \sca{\vt{S}_{n+1}}{\widebar{\vt{S}}_{n}} 
+ \sca{\vt{S}_{n}}{\widebar{\vt{S}}_{n+1}} + \sca{\vt{S}_{n}}{\widebar{\vt{S}}_{n-1}} + \sca{\vt{S}_{n-1}}{\widebar{\vt{S}}_{n}} 
\right)  \left( \vt{S}_{n+1}+\vt{S}_{n-1} \right), 
\nonumber \\
\label{dgvhYO-1} \\
\widebar{\vt{S}}_{n,\tau} = v_n v_{n+1} \left( \widebar{\vt{S}}_{n+2}+\widebar{\vt{S}}_{n} \right)
	- v_n v_{n-1} \left( \widebar{\vt{S}}_{n}+\widebar{\vt{S}}_{n-2} \right)
\nonumber \\ \hspace{11mm}
\mbox{} - \frac{\mathrm{i}}{2} v_{n} \left( \sca{\vt{S}_{n+1}}{\widebar{\vt{S}}_{n}} 
+ \sca{\vt{S}_{n}}{\widebar{\vt{S}}_{n+1}} + \sca{\vt{S}_{n}}{\widebar{\vt{S}}_{n-1}} + \sca{\vt{S}_{n-1}}{\widebar{\vt{S}}_{n}} 
\right) \left( \widebar{\vt{S}}_{n+1}+\widebar{\vt{S}}_{n-1} \right), 
\nonumber \\
\label{dgvhYO-2} \\
v_{n,\tau} = 2 v_n^2 \left( v_{n+1} - v_{n-1} \right) 
\nonumber \\ \hspace{11mm}
\mbox{} + \frac{\mathrm{i}}{2} v_n v_{n+1} 
\left( \sca{\vt{S}_{n+2}}{\widebar{\vt{S}}_{n}} - \sca{\vt{S}_{n}}{\widebar{\vt{S}}_{n+2}} \right)  -  \frac{\mathrm{i}}{2} v_n v_{n-1} 
\left( \sca{\vt{S}_{n}}{\widebar{\vt{S}}_{n-2}} - \sca{\vt{S}_{n-2}}{\widebar{\vt{S}}_{n}} \right) 
\nonumber \\ \hspace{11mm}
\mbox{} - \frac{1}{4} v_{n} \left( \sca{\vt{S}_{n+1}}{\widebar{\vt{S}}_{n}} 
+ \sca{\vt{S}_{n}}{\widebar{\vt{S}}_{n+1}} + \sca{\vt{S}_{n}}{\widebar{\vt{S}}_{n-1}} + \sca{\vt{S}_{n-1}}{\widebar{\vt{S}}_{n}} 
\right) 
\nonumber \\ \hspace{11mm}
\mbox{} \times
\left( \sca{\vt{S}_{n+1}}{\widebar{\vt{S}}_{n}} 
+ \sca{\vt{S}_{n}}{\widebar{\vt{S}}_{n+1}} - \sca{\vt{S}_{n}}{\widebar{\vt{S}}_{n-1}} - \sca{\vt{S}_{n-1}}{\widebar{\vt{S}}_{n}} 
\right). 
\label{dgvhYO-3} 
\end{subnumcases}
\end{proposition}

We leave it to the reader 
to verify directly that 
the two time evolutions indeed commute, i.e., 
\[
\frac{\partial^2 \vt{S}_{n}}{\partial_t \partial_\tau} = \frac{\partial^2 \vt{S}_{n}}{\partial_\tau \partial_t}, 
\hspace{5mm} 
\frac{\partial^2 \widebar{\vt{S}}_{n}}{\partial_t \partial_\tau} = \frac{\partial^2 \widebar{\vt{S}}_{n}}{\partial_\tau \partial_t}, 
\hspace{5mm} 
\frac{\partial^2 v_{n}}{\partial_t \partial_\tau} = \frac{\partial^2 v_{n}}{\partial_\tau \partial_t}. 
\]

In the absence of 
$\vt{S}_{n}$ and $ \widebar{\vt{S}}_{n}$, 
the higher discrete multicomponent Yajima--Oikawa system (\ref{dgvhYO}) 
reduces to a modified version of the Volterra lattice~\cite{Wadati76}: 
\begin{equation}
v_{n,\tau} = 2 v_n^2 \left( v_{n+1} - v_{n-1} \right), 
\label{mLV}
\end{equation}
which is related to the  Volterra lattice 
\begin{equation}
u_{n,\tau} = u_n \left( u_{n+1} - u_{n-1} \right), 
\nonumber
\end{equation}
by the transformation \mbox{$u_{n} := 2 v_n v_{n+1} $}. 
This implies that 
the discrete 
Yajima--Oikawa hierarchy 
generalizes 
a modified version of the Volterra lattice hierarchy in the same way 
as the continuous Yajima--Oikawa hierarchy generalizes the KdV hierarchy~\cite{Cheng92}. 

The
three-component spectral problem (\ref{YO-L}) for $\psi_n$, $\vt{\phi}_n$ and $\vt{\chi}_n$ 
can 
be 
rewritten as a nonlocal spectral problem for the single scalar 
component 
$\psi_n$. 
That is, from (\ref{YO-L2}) and (\ref{YO-L3}), we obtain 
\[
\vt{\phi}_n = \frac{\mathrm{i}}{2} \sum_{j=-\infty}^{n-1} \widebar{\vt{S}}_{j} \left( \psi_{j+1} + \psi_{j-1} \right),
\]
and 
\[
\vt{\chi}_n = \frac{\mathrm{i}}{2} \sum_{j=-\infty}^{n-1} (-1)^{n-j-1} \vt{S}_{j}  \left( \psi_{j+1} + \psi_{j-1} \right),
\]
under the assumption that 
$\vt{\phi}_n$ and $\vt{\chi}_n$ approach zero 
as \mbox{$n \to - \infty$}. Thus, we arrive at the following proposition. 
\begin{proposition}
\label{prop2.3}
If $\vt{S}_n$, $\widebar{\vt{S}}_{n} $ and $v_{n}$ satisfy 
the discrete multicomponent Yajima--Oikawa system \mbox{$(\ref{dgvYO})$}, 
the overdetermined linear equations 
for $\psi_n$: 
\begin{equation}
\label{YO-nonlocalL}
v_{n} \left( \psi_{n+1} + \psi_{n-1} \right) = \lambda \psi_{n} 
	- \frac{\mathrm{i}}{2} \sum_{j=-\infty}^{n-1} \left[ 
\sca{\vt{S}_n}{\widebar{\vt{S}}_{j}} + (-1)^{n-j-1} \sca{\vt{S}_{j} }{\widebar{\vt{S}}_{n}} \right] 
 \left( \psi_{j+1} + \psi_{j-1} \right), 
\end{equation}
and 
\begin{equation}
\label{YO-nonlocalM}
\mathrm{i} \psi_{n,t} = v_{n} \left( \psi_{n+1} + \psi_{n-1} \right) - c \psi_n, 
\end{equation}
are compatible, 
where $\lambda$ is 
a constant spectral parameter. 
\end{proposition}

A 
straightforward 
calculation shows that 
the compatibility condition 
for 
(\ref{YO-nonlocalL}) and (\ref{YO-nonlocalM}) 
is given by
\begin{align}
& \mathrm{i} \left[ 
v_{n,t} - \frac{1}{2} v_n \hspace{1pt} \boldsymbol{\Delta}_n^+ 
\left( \sca{\vt{S}_{n}}{\widebar{\vt{S}}_{n-1}} + \sca{\vt{S}_{n-1}}{\widebar{\vt{S}}_{n}} \right)
\right] \left( \psi_{n+1} + \psi_{n-1} \right)
\nonumber \\
& \mbox{} + \frac{\mathrm{i}}{2} \sum_{j=-\infty}^{n-1} \left[ \sca{\mathrm{i} \vt{S}_{n,t}
	 - v_n \left( \vt{S}_{n+1}+\vt{S}_{n-1} \right)}{\widebar{\vt{S}}_{j}} 
+ \sca{\vt{S}_{n}}{\mathrm{i} \widebar{\vt{S}}_{j,t} + v_j \left( \widebar{\vt{S}}_{j+1}+\widebar{\vt{S}}_{j-1} \right)} \right. 
\nonumber \\
& \left. \mbox{} + (-1)^{n-j-1} \sca{\mathrm{i} \vt{S}_{j,t}
	 - v_j \left( \vt{S}_{j+1}+\vt{S}_{j-1} \right)}{\widebar{\vt{S}}_{n}} 
+ (-1)^{n-j-1} \sca{\vt{S}_{j}}{\mathrm{i} \widebar{\vt{S}}_{n,t} + v_n \left( \widebar{\vt{S}}_{n+1}+\widebar{\vt{S}}_{n-1} \right)} \right] 
\nonumber \\
& \mbox{} \times \left( \psi_{j+1} + \psi_{j-1} \right) = 0, 
\nonumber
\end{align}
which 
does not contain the arbitrary constant $c$ 
and 
is equivalent to the set of relations: 
\begin{align}
& 
v_{n,t} = \frac{1}{2} v_n \hspace{1pt} \boldsymbol{\Delta}_n^+ 
\left( \sca{\vt{S}_{n}}{\widebar{\vt{S}}_{n-1}} + \sca{\vt{S}_{n-1}}{\widebar{\vt{S}}_{n}} \right), 
\nonumber \\[2mm]
& \sca{\mathrm{i} \vt{S}_{n,t}
	 - v_n \left( \vt{S}_{n+1}+\vt{S}_{n-1} \right)}{\widebar{\vt{S}}_{j}} 
+ \sca{\vt{S}_{n}}{\mathrm{i} \widebar{\vt{S}}_{j,t} + v_j \left( \widebar{\vt{S}}_{j+1}+\widebar{\vt{S}}_{j-1} \right)} 
\nonumber \\
& \mbox{} + (-1)^{n-j-1} \sca{\mathrm{i} \vt{S}_{j,t}
	 - v_j \left( \vt{S}_{j+1}+\vt{S}_{j-1} \right)}{\widebar{\vt{S}}_{n}} 
+ (-1)^{n-j-1} \sca{\vt{S}_{j}}{\mathrm{i} \widebar{\vt{S}}_{n,t} + v_n \left( \widebar{\vt{S}}_{n+1}+\widebar{\vt{S}}_{n-1} \right)} 
\nonumber \\
& = 0, \hspace{5mm} j \le n-1. 
\nonumber
\end{align}
Thus, 
(\ref{dgvYO}) 
implies 
the compatibility of (\ref{YO-nonlocalL}) and (\ref{YO-nonlocalM}), 
but {\em not} vice versa. In short, 
the scalar nonlocal spectral problem (\ref{YO-nonlocalL}) and 
the isospectral time-evolution equation (\ref{YO-nonlocalM}) 
provide a {\em weak} Lax-pair representation for the discrete multicomponent 
Yajima--Oikawa system (\ref{dgvYO}). 
%
This is evident from the fact that the scalar 
nonlocal spectral problem 
(\ref{YO-nonlocalL}) is invariant under the transformation 
\[
\vt{S}_n \to g \vt{S}_n, \hspace{5mm} \widebar{\vt{S}}_{n} \to g^{-1} \widebar{\vt{S}}_{n}, 
\]
where $g$ is a nonzero $n$-independent scalar. 
In particular, 
the spatial Lax operator defined by 
(\ref{YO-nonlocalL}) is invariant 
under the time evolution corresponding to the zeroth flow of the hierarchy: 
\[
v_n (t_0)= v_n (0), \hspace{5mm}
\vt{S}_n (t_0)=
\exp \left( \mathrm{i} d t_0 \right) \vt{S}_n (0), 
	\hspace{5mm} \widebar{\vt{S}}_{n} (t_0) = 
\exp \left( -\mathrm{i} d t_0 \right) \widebar{\vt{S}}_{n} (0), 
\]
so 
the value of $c$ in the discrete multicomponent Yajima--Oikawa 
system (\ref{dgvYO}) can no longer be 
determined from 
the compatibility of (\ref{YO-nonlocalL}) and (\ref{YO-nonlocalM}). 
This is 
an intrinsic drawback 
of the scalar nonlocal Lax-pair representations 
for 
Yajima--Oikawa-type systems. 

To 
consider the 
continuous limit, 
we set 
\begin{align}
v_n (t) = \frac{1}{\varDelta^2} + \frac{1}{2} L (n \varDelta, t), 
\hspace{5mm} 
\vt{S}_n(t) = \vt{S}(n \varDelta, t), 
\hspace{5mm} 
\widebar{\vt{S}}_n(t) = \varDelta \hspace{1pt} \widebar{\vt{S}}(n \varDelta, t), 
\nonumber 
\end{align}
and 
\begin{align}
x = 
n \varDelta, \hspace{5mm} c= \frac{2}{\varDelta^2}, 
\nonumber 
\end{align}
where $\varDelta$ is a lattice parameter. 
Then, in the continuous limit \mbox{$\varDelta \to 0$}, 
the  discrete multicomponent Yajima--Oikawa system (\ref{dgvYO}) 
reduces to 
\begin{subnumcases}{\label{gvYO}}
\mathrm{i} \vt{S}_{t} =  \vt{S}_{xx} + L \vt{S}, 
\label{gvYO-1} \\[1pt]
\mathrm{i} \widebar{\vt{S}}_{t} = - \widebar{\vt{S}}_{xx} - L \widebar{\vt{S}}, 
\label{gvYO-2} \\[1pt]
L_{t} = 2 \sca{\vt{S}}{\widebar{\vt{S}}}_x. 
\label{gvYO-3} 
\end{subnumcases}
%
Thus, 
we obtain 
the continuous multicomponent Yajima--Oikawa system (\ref{vYO}) 
by further 
setting 
\mbox{$\widebar{\vt{S}}=\Sigma \vt{S}^\ast$}. 
The Lax-pair representation for (\ref{gvYO}) is 
obtained by setting 
\[
\psi_n(t) = \psi (n \varDelta, t), 
\hspace{5mm} 
\vt{\phi}_n(t) = \vt{\phi} (n \varDelta, t), 
\hspace{5mm}
\lambda= \frac{2}{\varDelta^2} +\zeta,
\]
and considering 
the continuous limit \mbox{$\varDelta \to 0$} of Proposition~\ref{prop2.1}; 
the third component $\vt{\chi}_n$ 
in the linear eigenfunction 
can be discarded in the 
continuous limit.  
%
\begin{proposition}
\label{prop2.4}
The 
nonreduced 
multicomponent Yajima--Oikawa system \mbox{$(\ref{gvYO})$} 
is equivalent to the 
compatibility condition 
of 
the overdetermined linear equations 
for $\psi$ and $\vt{\phi}$: 
\[
\left\{
\begin{split}
& \psi_{xx} + L \psi = \zeta \psi 
	-\sca{\vt{S}}{\vt{\phi}},
\nonumber 
\\[1pt]
& \vt{\phi}_{x} = \mathrm{i} \widebar{\vt{S}} \psi,
\nonumber 
\nonumber
\end{split}
\right.
\]
and 
\[
\left\{
\begin{split}
& \mathrm{i} \psi_{t} = \psi_{xx} + L \psi, 
\nonumber \\[1pt]
& \vt{\phi}_{t} = \widebar{\vt{S}} \psi_{x} - \widebar{\vt{S}}_{x} \psi, 
\nonumber
\end{split}
\right.
\]
where $\zeta$ is 
a constant spectral parameter. 
\end{proposition}

A weak Lax-pair representation for (\ref{gvYO}) 
is obtained by rewriting $\vt{\phi}$ as
\[
\vt{\phi}(x,t) = \mathrm{i} \int^x \widebar{\vt{S}} (y,t) \psi (y,t) \mathrm{d} y. 
\]
\begin{proposition}
If $\vt{S}$, $\widebar{\vt{S}}$ and $L$ satisfy 
the nonreduced 
multicomponent Yajima--Oikawa system 
\mbox{$(\ref{gvYO})$}, 
then 
the 
overdetermined linear equations for $\psi$: 
\begin{equation}
\psi_{xx} + L \psi + \mathrm{i} \sca{\vt{S}}{\partial_x^{-1} \left( \widebar{\vt{S}} \psi \right)} = \zeta \psi,
\end{equation}
and 
\begin{equation}
\mathrm{i} \psi_{t} = \psi_{xx} + L \psi, 
\end{equation}
are compatible but not vice versa, 
where $\zeta$ is 
a constant spectral parameter. 
\end{proposition}

\section{Conservation laws}

In this section, we show that the discrete multicomponent Yajima--Oikawa system (\ref{dgvYO}) 
possesses an infinite number of conservation laws, which can be constructed 
from its Lax-pair representation 
in a recursive manner. As is clear from the following derivation, 
the 
conserved densities are determined 
from 
the spectral problem (\ref{YO-L}) only, 
so the higher flows of the  discrete multicomponent Yajima--Oikawa hierarchy 
such as 
(\ref{dgvhYO}) 
have the same set of 
conserved quantities 
as (\ref{dgvYO}). 

We 
start with 
the trivial identity:
\begin{equation}
\left[ \log \left( \frac{\psi_n}{\psi_{n-1}} \right) \right]_t 
=\boldsymbol{\Delta}_n^+ \left[ \frac{\psi_{n-1,t}}{\psi_{n-1}} \right], 
\label{start_point}
\end{equation}
and use 
the quantity \mbox{$\log \left( \psi_n/\psi_{n-1}\right)= \log \psi_n - \log \psi_{n-1}$} 
as a generating function of the 
conserved densities~\cite{Wadati1977,Haberman}. 
We set 
\begin{subnumcases}{\label{FGH_def}}
\frac{\psi_n}{\psi_{n-1}} =: \frac{1}{\lambda} v_n F_n, 
\label{F_def}
\\[1pt]
\frac{\vt{\phi}_n}{\psi_{n-1}}=: - \frac{\mathrm{i}}{2} \widebar{\vt{S}}_n + \frac{1}{\lambda} v_n \vt{G}_n, 
\label{G_def}
\\[1pt]
\frac{\vt{\chi}_n}{\psi_{n-1}}=: \frac{\mathrm{i}}{2} \vt{S}_n + \frac{1}{\lambda} v_n \vt{H}_n,
\label{H_def}
\end{subnumcases}
where $F_n$ is a scalar and $\vt{G}_n$ and $\vt{H}_n$ are column vectors, 
and rewrite the spectral problem (\ref{YO-L}) as 
\begin{subnumcases}{\label{YO-R}}
F_n = 1+ \frac{1}{\lambda^2} v_n v_{n+1} F_n F_{n+1} 
	+ \frac{1}{\lambda} \sca{\vt{S}_n}{\vt{G}_n} + \frac{1}{\lambda} \sca{\vt{H}_n}{\widebar{\vt{S}}_{n}}, 
\label{YO-R1}
\\[1pt]
\vt{G}_n = F_n \left(  - \frac{\mathrm{i}}{2} \widebar{\vt{S}}_{n+1} + \frac{1}{\lambda} v_{n+1} \vt{G}_{n+1} 
\right) - \frac{\mathrm{i}}{2\lambda} v_{n+1} F_n F_{n+1} \widebar{\vt{S}}_{n},
\label{YO-R2}
\\[1pt]
\vt{H}_n = -F_n \left(\frac{\mathrm{i}}{2} \vt{S}_{n+1} + \frac{1}{\lambda} v_{n+1} \vt{H}_{n+1} \right)
+ \frac{\mathrm{i}}{2 \lambda} v_{n+1} F_n F_{n+1}  \vt{S}_{n}.
\label{YO-R3}
\end{subnumcases}
The trivial identity (\ref{start_point}) can be rewritten 
with the aid of (\ref{F_def}) and (\ref{YO-M1}) as 
a nontrivial conservation law: 
\begin{equation}
\left[ \log v_n + \log F_n \right]_t 
= 
\boldsymbol{\Delta}_n^+ \left[ -\mathrm{i} \left( \frac{1}{\lambda} v_{n-1} v_{n} F_{n} + \lambda \frac{1}{F_{n-1}} \right) \right]. 
\label{generating}
\end{equation}
%
The set of relations (\ref{YO-R}) 
allows us to express 
the three quantities 
 $F_n$, $\vt{G}_n$ and $\vt{H}_n$ 
as power series 
in $1/\lambda$:
\begin{align}
F_n = \sum_{j=0}^\infty \frac{1}{\lambda^j} F_n^{(j)}, 
\hspace{5mm}
\vt{G}_n =  \sum_{j=0}^\infty \frac{1}{\lambda^j} \vt{G}_n^{(j)}, 
\hspace{5mm}
\vt{H}_n = \sum_{j=0}^\infty \frac{1}{\lambda^j} \vt{H}_n^{(j)},
\label{FGH_exp}
\end{align}
%
where the 
coefficients 
$\{ F_n^{(j)} \}_{j \ge 0}$, $\{ \vt{G}_n^{(j)} \}_{j \ge 0}$ and $\{ \vt{H}_n^{(j)} \}_{j \ge 0}$ 
are local functions of the dependent variables $v_n$, $\vt{S}_n$ and $\widebar{\vt{S}}_n$. 
Substituting (\ref{FGH_exp}) into (\ref{YO-R}), we obtain the recurrence 
relations for the coefficients 
$\{ F_n^{(j)} \}_{j \ge 0}$, $\{ \vt{G}_n^{(j)} \}_{j \ge 0}$ and $\{ \vt{H}_n^{(j)} \}_{j \ge 0}$: 
\begin{subnumcases}{\label{YO-Re}}
F_n^{(j)} = 
v_n v_{n+1} \sum_{k=0}^{j-2} F_n^{(k)} F_{n+1}^{(j-k-2)}
	+ \sca{\vt{S}_n}{\vt{G}_n^{(j-1)}} + \sca{\vt{H}_n^{(j-1)}}{\widebar{\vt{S}}_{n}}, 
\;\;\;  j \ge 1, 
\nonumber \\
\label{YO-Re1}
\\
\vt{G}_n^{(j)} = - \frac{\mathrm{i}}{2}  F_n^{(j)} \widebar{\vt{S}}_{n+1} 
 + v_{n+1} \sum_{k=0}^{j-1} F_n^{(k)} \vt{G}_{n+1}^{(j-k-1)} 
 - \frac{\mathrm{i}}{2} v_{n+1} \sum_{k=0}^{j-1} F_n^{(k)} F_{n+1}^{(j-k-1)} \widebar{\vt{S}}_{n},
\nonumber \\
\label{YO-Re2}
\\
\vt{H}_n^{(j)} = - \frac{\mathrm{i}}{2} F_n^{(j)} \vt{S}_{n+1} 
 - v_{n+1} \sum_{k=0}^{j-1} F_n^{(k)} \vt{H}_{n+1}^{(j-k-1)}  
+ \frac{\mathrm{i}}{2} v_{n+1} \sum_{k=0}^{j-1} F_n^{(k)} F_{n+1}^{(j-k-1)} \vt{S}_{n},
\nonumber \\
\label{YO-Re3}
\end{subnumcases}
where \mbox{$F_n^{(0)}=1$}. 
Explicit expressions for 
the first few 
coefficients are
as follows: 
\begin{align}
F_n &= 1+ \frac{1}{\lambda} \left(  - \frac{\mathrm{i}}{2} \sca{\vt{S}_{n+1}}{\widebar{\vt{S}}_{n}} 
	- \frac{\mathrm{i}}{2} \sca{\vt{S}_{n}}{\widebar{\vt{S}}_{n+1}} \right) 
\nonumber \\ 
& \hphantom{=} \; \mbox{}+\frac{1}{\lambda^2} \left[ v_n v_{n+1} 
	+\frac{\mathrm{i}}{2} v_{n+1} \left( \sca{\vt{S}_{n+2}}{\widebar{\vt{S}}_{n}} 
	- \sca{\vt{S}_{n}}{\widebar{\vt{S}}_{n+2}} \right) 
	-\frac{1}{4}  \left( \sca{\vt{S}_{n+1}}{\widebar{\vt{S}}_{n}} + \sca{\vt{S}_{n}}{\widebar{\vt{S}}_{n+1}} \right)^2 \right]
\nonumber \\ 
& \hphantom{=} \; \mbox{}+	O\left( \frac{1}{\lambda^3} \right), 
\nonumber
\\[2pt]
\vt{G}_n &=  - \frac{\mathrm{i}}{2} \widebar{\vt{S}}_{n+1} +\frac{1}{\lambda} 
\left[ - \frac{\mathrm{i}}{2} v_{n+1} \left( \widebar{\vt{S}}_{n+2} + \widebar{\vt{S}}_{n} \right)  
-\frac{1}{4} \left( \sca{\vt{S}_{n+1}}{\widebar{\vt{S}}_{n}} + \sca{\vt{S}_{n}}{\widebar{\vt{S}}_{n+1}} \right) \widebar{\vt{S}}_{n+1} 
\right]
\nonumber \\ 
\nonumber 
& \hphantom{=} \; \mbox{}+	O\left( \frac{1}{\lambda^2} \right), 
\\[2pt]
\vt{H}_n &= -\frac{\mathrm{i}}{2} \vt{S}_{n+1} + \frac{1}{\lambda}
\left[ \frac{\mathrm{i}}{2} v_{n+1} \left( \vt{S}_{n+2} + \vt{S}_{n} \right)  
-\frac{1}{4} \left( \sca{\vt{S}_{n+1}}{\widebar{\vt{S}}_{n}} + \sca{\vt{S}_{n}}{\widebar{\vt{S}}_{n+1}} \right) \vt{S}_{n+1} 
\right]
\nonumber \\ 
& \hphantom{=} \; \mbox{}+	O\left( \frac{1}{\lambda^2} \right). 
\nonumber
\end{align}

Now, we can state the main result of this section. 
\begin{proposition}
\label{prop3.1}
Let 
$\{ F_n^{(j)} \}_{j \ge 0}$
together with $\{ \vt{G}_n^{(j)} \}_{j \ge 0}$ and $\{ \vt{H}_n^{(j)} \}_{j \ge 0}$ 
be defined 
by the recurrence 
relations \mbox{$(\ref{YO-Re})$} with \mbox{$F_n^{(0)}=1$}. 
Then, the 
discrete multicomponent Yajima--Oikawa system \mbox{$(\ref{dgvYO})$} 
possesses
an infinite number of conservation laws, which can be obtained by 
comparing the coefficients of 
different 
powers of $1/\lambda$ 
on both sides of the equality: 
\begin{align}
& \left[ \log v_n + \log \left( 1+\sum_{j=1}^\infty \frac{1}{\lambda^j} F_n^{(j)} \right) \right]_t 
\nonumber \\[2pt]
& = 
\boldsymbol{\Delta}_n^+ \left[ - \frac{\mathrm{i}}{\lambda} v_{n-1} v_{n} 
	\left( 1+\sum_{j=1}^\infty \frac{1}{\lambda^j} F_n^{(j)} \right)
 -\mathrm{i}  \lambda \frac{1}{1+\sum_{j=1}^\infty \frac{1}{\lambda^j} F_{n-1}^{(j)}} \right]. 
\label{con_ge}
\end{align}
\end{proposition}

Noting that 
\begin{align}
& \log \left( 1+x \right) = x - \frac{x^2}{2} + \frac{x^3}{3} 
-\cdots,
\nonumber \\[1pt]
& \frac{1}{1+x} = 1 -x +x^2 -x^3 + \cdots, 
\nonumber 
\end{align}
the first four conservation laws obtained 
from (\ref{con_ge}) are as follows: 
\begin{align}
& \left( \log v_n \right)_t 
 = \boldsymbol{\Delta}_n^+ \left[ \frac{1}{2} \left( \sca{\vt{S}_{n}}{\widebar{\vt{S}}_{n-1}}  
	+ \sca{\vt{S}_{n-1}}{\widebar{\vt{S}}_{n}} \right) \right], 
\nonumber \\[6pt]
& \left( \sca{\vt{S}_{n+1}}{\widebar{\vt{S}}_{n}}  
	+ \sca{\vt{S}_{n}}{\widebar{\vt{S}}_{n+1}} \right)_t 
 = \boldsymbol{\Delta}_n^+ \left[ - \mathrm{i} v_{n} \left( \sca{\vt{S}_{n+1}}{\widebar{\vt{S}}_{n-1}} 
	- \sca{\vt{S}_{n-1}}{\widebar{\vt{S}}_{n+1}} \right) \right], 
\nonumber \\[6pt]
& \left[ v_n v_{n+1} + \frac{\mathrm{i}}{2} v_{n+1} \left( \sca{\vt{S}_{n+2}}{\widebar{\vt{S}}_{n}} 
	- \sca{\vt{S}_{n}}{\widebar{\vt{S}}_{n+2}} \right) -\frac{1}{8} \left( \sca{\vt{S}_{n+1}}{\widebar{\vt{S}}_{n}}  
	+ \sca{\vt{S}_{n}}{\widebar{\vt{S}}_{n+1}} \right)^2 \right]_t 
\nonumber \\
& = \boldsymbol{\Delta}_n^+ \left[ \frac{1}{2} v_n v_{n+1} \left( \sca{\vt{S}_{n+2}}{\widebar{\vt{S}}_{n-1}}  
	+ \sca{\vt{S}_{n-1}}{\widebar{\vt{S}}_{n+2}} + \sca{\vt{S}_{n}}{\widebar{\vt{S}}_{n-1}}  
	+ \sca{\vt{S}_{n-1}}{\widebar{\vt{S}}_{n}} \right) \right. 
\nonumber \\ 
& \hphantom{=} 
\left. \mbox{} + \frac{\mathrm{i}}{4} v_n
	\left( \sca{\vt{S}_{n+1}}{\widebar{\vt{S}}_{n}}  
	+ \sca{\vt{S}_{n}}{\widebar{\vt{S}}_{n+1}} \right) \left( \sca{\vt{S}_{n+1}}{\widebar{\vt{S}}_{n-1}}  
	- \sca{\vt{S}_{n-1}}{\widebar{\vt{S}}_{n+1}} \right) \right],
\nonumber \\[6pt]
& \left[  v_n v_{n+1} \left( \sca{\vt{S}_{n+2}}{\widebar{\vt{S}}_{n+1}}  
	+ \sca{\vt{S}_{n+1}}{\widebar{\vt{S}}_{n+2}} + \sca{\vt{S}_{n+1}}{\widebar{\vt{S}}_{n}}  
	+ \sca{\vt{S}_{n}}{\widebar{\vt{S}}_{n+1}} \right)
\right. 
\nonumber \\
& \; \mbox{} + v_{n+1} v_{n+2} \left( \sca{\vt{S}_{n+3}}{\widebar{\vt{S}}_{n}}  
	+ \sca{\vt{S}_{n}}{\widebar{\vt{S}}_{n+3}} + \sca{\vt{S}_{n+1}}{\widebar{\vt{S}}_{n}}  
	+ \sca{\vt{S}_{n}}{\widebar{\vt{S}}_{n+1}} \right)
\nonumber \\
& \; \mbox{} +\frac{\mathrm{i}}{2} v_{n+1}  \left( \sca{\vt{S}_{n+2}}{\widebar{\vt{S}}_{n+1}}  
	+ \sca{\vt{S}_{n+1}}{\widebar{\vt{S}}_{n+2}} + \sca{\vt{S}_{n+1}}{\widebar{\vt{S}}_{n}}  
	+ \sca{\vt{S}_{n}}{\widebar{\vt{S}}_{n+1}} \right) 
\nonumber \\
& \; \left. \mbox{} \times 
\left( \sca{\vt{S}_{n+2}}{\widebar{\vt{S}}_{n}}  
	- \sca{\vt{S}_{n}}{\widebar{\vt{S}}_{n+2}} \right) 
- \frac{1}{12} \left( \sca{\vt{S}_{n+1}}{\widebar{\vt{S}}_{n}}  
	+ \sca{\vt{S}_{n}}{\widebar{\vt{S}}_{n+1}} \right)^3 
\right]_t
\nonumber \\
& = \boldsymbol{\Delta}_n^+ \left[ \ldots \right]. 
\nonumber
\end{align}
Here, 
the 
flux in the fourth conservation law is 
lengthy and omitted. 

Not all 
of the 
conservation laws of 
the discrete multicomponent Yajima--Oikawa system (\ref{dgvYO}) 
can be obtained in this way. 
First, 
the second conservation law can be generalized to 
the componentwise form: 
\[
 \left( S_{n+1}^{(j)} \widebar{S}_{n}^{(k)}  
	+ S_{n}^{(j)} \widebar{S}_{n+1}^{(k)} \right)_t 
 = \boldsymbol{\Delta}_n^+ \left[ - \mathrm{i} v_{n} \left( S_{n+1}^{(j)} \widebar{S}_{n-1}^{(k)} 
	- S_{n-1}^{(j)} \widebar{S}_{n+1}^{(k)} \right) \right],
\]
where $S_{n}^{(j)}$ is the $j$th component of the column vector $\vt{S}_n$ and 
$\widebar{S}_{n}^{(k)}$ is the $k$th component of the column vector $\widebar{\vt{S}}_{n}$. 
Second, in the case where $\vt{S}_n$ and $\widebar{\vt{S}}_{n}$ are 
scalar, namely,  \mbox{$\vt{S}_n =S_n$} and $\widebar{\vt{S}}_{n}= \widebar{S}_n$, 
(\ref{dgvYO}) admits 
one more conservation law: 
\begin{align}
& \left\{ v_n \left( S_{n+1} + S_{n-1} \right) 
	\left( \widebar{S}_{n+1} + \widebar{S}_{n-1} \right) 
	+ \frac{\mathrm{i}}{4} \left[ \left(S_{n+1} \widebar{S}_{n}\right)^2 - \left( S_{n} \widebar{S}_{n+1}\right)^2 \right] 
	\right\}_t 
\nonumber \\
& = \boldsymbol{\Delta}_n^+ \left\{ -\mathrm{i} v_n v_{n-1} \left[ \left( S_{n+1} + S_{n-1} \right) 
	\left( \widebar{S}_{n} + \widebar{S}_{n-2} \right) -  \left( S_{n} + S_{n-2} \right) 
	\left( \widebar{S}_{n+1} + \widebar{S}_{n-1} \right) \right] \right.
\nonumber \\
& \hphantom{=} 
\left. \mbox{} + \frac{1}{2} v_n \left[ S_{n-1} \widebar{S}_{n-1} 
\left( S_{n} \widebar{S}_{n-1} + S_{n-1} \widebar{S}_{n} \right) +  S_{n+1} \widebar{S}_{n-1}  S_{n} \widebar{S}_{n-1} 
+  S_{n-1} \widebar{S}_{n+1}  S_{n-1} \widebar{S}_{n} 
\right] 
\right\},
\nonumber 
\end{align}
which can be found with the aid of 
the Mathematica package 
``diffdens.m''~\cite{Hereman98}
available from
\\
\hspace{10mm}
https://inside.mines.edu/\verb|~|whereman/software/diffdens/  .

\section{Relations to 
discrete nonlinear Schr\"odinger hierarchies}

In this section, 
we show 
how 
the discrete multicomponent Yajima--Oikawa system (\ref{dgvYO}) 
can be related to 
two 
integrable discrete nonlinear Schr\"odinger hierarchies:
the Ablowitz--Ladik 
hierarchy~\cite{AL1} 
and 
the Konopelchenko--Chudnovsky hierarchy~\cite{Kono82,Chud1,Chud2}.

\subsection{Ablowitz--Ladik hierarchy}
By changing the dependent variables as 
\mbox{$\vt{S}_n= \alpha^{n} \vt{q}_n$} and \mbox{$\widebar{\vt{S}}_n= \alpha^{-n} \widebar{\vt{q}}_n$} 
with a nonzero constant $\alpha$, 
the discrete multicomponent Yajima--Oikawa system (\ref{dgvYO}) 
is transformed to 
the following form: 
\begin{subnumcases}{\label{t_dgvYO}}
\mathrm{i} \vt{q}_{n,t} = v_n \left( \alpha \vt{q}_{n+1}+ \frac{1}{\alpha} \vt{q}_{n-1} \right) -c \vt{q}_{n}, 
\label{t_dgvYO-1} \\[1pt]
\mathrm{i} \widebar{\vt{q}}_{n,t} = -v_n \left( \frac{1}{\alpha} \widebar{\vt{q}}_{n+1}+ \alpha \widebar{\vt{q}}_{n-1} \right)
	+ c \widebar{\vt{q}}_{n}, 
\label{t_dgvYO-2} \\[1pt]
v_{n,t} = \frac{1}{2} v_n \hspace{1pt} \boldsymbol{\Delta}_n^+ 
\left( \alpha \sca{\vt{q}_{n}}{\widebar{\vt{q}}_{n-1}} + \frac{1}{\alpha} \sca{\vt{q}_{n-1}}{\widebar{\vt{q}}_{n}} \right). 
\label{t_dgvYO-3} 
\end{subnumcases}
%
Thus, 
by rescaling the time variable as \mbox{$\partial_t =: \alpha \partial_{t_1}$}and taking the limit 
\mbox{$\alpha \to \infty$}, 
(\ref{t_dgvYO}) reduces to 
\begin{subnumcases}{\label{tl1_dgvYO}}
\mathrm{i} \vt{q}_{n,t_1} = v_n \vt{q}_{n+1}, 
\label{tl1_dgvYO-1} \\[1pt]
\mathrm{i} \widebar{\vt{q}}_{n,t_1} = -v_n \widebar{\vt{q}}_{n-1}, 
\label{tl1_dgvYO-2} \\[1pt]
v_{n,t_1} = \frac{1}{2} v_n \hspace{1pt} \boldsymbol{\Delta}_n^+ 
\left( \sca{\vt{q}_{n}}{\widebar{\vt{q}}_{n-1}} \right), 
\label{tl1_dgvYO-3} 
\end{subnumcases}
and by rescaling the time variable as \mbox{$\partial_t =: \frac{1}{\alpha} \partial_{t_{-1}}$}and taking the limit 
\mbox{$\alpha \to 0$}, (\ref{t_dgvYO}) reduces to 
\begin{subnumcases}{\label{tl2_dgvYO}}
\mathrm{i} \vt{q}_{n,t_{-1}} = v_n \vt{q}_{n-1}, 
\label{tl2_dgvYO-1} \\[1pt]
\mathrm{i} \widebar{\vt{q}}_{n,t_{-1}} = -v_n \widebar{\vt{q}}_{n+1}, 
\label{tl2_dgvYO-2} \\[1pt]
v_{n,t_{-1}} = \frac{1}{2} v_n \hspace{1pt} \boldsymbol{\Delta}_n^+ \left( \sca{\vt{q}_{n-1}}{\widebar{\vt{q}}_{n}} \right). 
\label{tl2_dgvYO-3} 
\end{subnumcases}
%
The two systems (\ref{tl1_dgvYO}) and (\ref{tl2_dgvYO}) 
are essentially 
equivalent and already known in the case of scalar $\vt{q}_{n}$ and $\widebar{\vt{q}}_{n}$
(see, {{\em e.g.}, (4.19) or (4.21) in \cite{Li2016}). 
Note that $\partial_{t_1}$ and $\partial_{t_{-1}}$ 
do not 
commute. 
%

From (\ref{tl1_dgvYO}), we have 
the relation
\[
\left( v_n - \frac{\mathrm{i}}{2}\sca{\vt{q}_{n}}{\widebar{\vt{q}}_n} \right)_{t_1} =0,
\]
which implies that 
\begin{equation}
v_n =k_n +  \frac{\mathrm{i}}{2}\sca{\vt{q}_{n}}{\widebar{\vt{q}}_n}, 
\nonumber 
\end{equation}
where $k_n$ is a ``constant'' independent of time $t_1$. 
Thus, 
(\ref{tl1_dgvYO}) reduces 
to 
the 
simpler form: 
\begin{subnumcases}{\label{tl3_dgvYO}}
\mathrm{i} \vt{q}_{n,t_1} = \left( k_n +  \frac{\mathrm{i}}{2}\sca{\vt{q}_{n}}{\widebar{\vt{q}}_n} \right) \vt{q}_{n+1}, 
\label{tl3_dgvYO-1} \\[1pt]
\mathrm{i} \widebar{\vt{q}}_{n,t_1} = - \left( k_n +  \frac{\mathrm{i}}{2}\sca{\vt{q}_{n}}{\widebar{\vt{q}}_n} \right) 
	\widebar{\vt{q}}_{n-1}. 
\label{tl3_dgvYO-2} 
\end{subnumcases}
%
For nonzero values of $k_n$, 
the change of variables 
\begin{equation} 
\vt{q}_{n} =  \left( \prod_{j
}^{n-1} \frac{1}{k_j} \right)
	\vt{a}_{n}, 
\hspace{5mm} 
  \frac{\mathrm{i}}{2} \widebar{\vt{q}}_{n} = \left( \prod_{j
}^{n} k_j \right) \vt{b}_{n}, 
\nonumber  
\end{equation}
can be used to normalize 
$k_n$ to $1$ and to remove $\frac{\mathrm{i}}{2}$ in (\ref{tl3_dgvYO}).
Thus, we 
obtain 
the vector generalization~\cite{GI2} of 
an elementary 
flow of 
the Ablowitz--Ladik hierarchy~\cite{AL1
}:
\begin{equation} 
\nonumber
\left\{ 
\begin{split}
& \mathrm{i} \vt{a}_{n,t_1} = \left( 1 + \sca{\vt{a}_{n}}{\vt{b}_n} \right) \vt{a}_{n+1}, 
\\[2pt]
& \mathrm{i} \vt{b}_{n,t_1} =- \left( 1 + \sca{\vt{a}_{n}}{\vt{b}_n} \right) \vt{b}_{n-1}. 
\end{split} 
\right. 
\end{equation}

\subsection{Konopelchenko--Chudnovsky hierarchy}

Under the 
boundary condition 
\mbox{$\lim_{n \to - \infty} \left( \sca{\vt{S}_{n}}{\widebar{\vt{S}}_{n-1}} + \sca{\vt{S}_{n-1}}{\widebar{\vt{S}}_{n}} \right) =0$},
(\ref{dgvYO-3}) implies the relation: 
\begin{equation}
\left( \prod_{j=-\infty}^n v_j \right)_t = \left( \prod_{j=-\infty}^n v_j \right) 
 \frac{1}{2} \left( \sca{\vt{S}_{n+1}}{\widebar{\vt{S}}_{n}} + \sca{\vt{S}_{n}}{\widebar{\vt{S}}_{n+1}} \right). 
\label{v_prod}
\end{equation}
Thus, the transformation of dependent variables: 
\begin{equation}
\vt{S}_n = \vt{r}_n \left( \prod_{j=-\infty}^n v_j \right)^{-\epsilon}, \hspace{5mm} 
\widebar{\vt{S}}_n = \widebar{\vt{r}}_n \left( \prod_{j=-\infty}^{n-1} v_j \right)^{\epsilon},
\label{trans_S_r}
\end{equation}
preserves the locality of the equations of motion, 
where $\epsilon$ is an arbitrary constant. 
In particular, the transformation (\ref{trans_S_r}) with \mbox{$\epsilon = -1$} 
changes the discrete multicomponent Yajima--Oikawa system (\ref{dgvYO}) 
to the following form: 
\begin{subnumcases}{\label{trans_dgvYO}}
\mathrm{i} \vt{r}_{n,t} = v_n v_{n+1} \vt{r}_{n+1} + \vt{r}_{n-1} 
	-\frac{\mathrm{i}}{2} \vt{r}_n \left( v_n v_{n+1} \sca{\vt{r}_{n+1}}{\widebar{\vt{r}}_{n}} 
	+ \sca{\vt{r}_{n}}{\widebar{\vt{r}}_{n+1}} \right) -c \vt{r}_{n}, \hspace{11mm}
\label{trans_dgvYO-1} \\[1pt]
\mathrm{i} \widebar{\vt{r}}_{n,t} = -\widebar{\vt{r}}_{n+1} -v_n v_{n-1} \widebar{\vt{r}}_{n-1} 
	+\frac{\mathrm{i}}{2} \widebar{\vt{r}}_n \left( v_n v_{n-1} \sca{\vt{r}_{n}}{\widebar{\vt{r}}_{n-1}} 
	+ \sca{\vt{r}_{n-1}}{\widebar{\vt{r}}_{n}} \right)
	+ c \widebar{\vt{r}}_{n}, \hspace{11mm}
\label{trans_dgvYO-2} \\[1pt]
v_{n,t} = \frac{1}{2} v_n \hspace{1pt} \boldsymbol{\Delta}_n^+ 
\left( v_n v_{n-1} \sca{\vt{r}_{n}}{\widebar{\vt{r}}_{n-1}} + \sca{\vt{r}_{n-1}}{\widebar{\vt{r}}_{n}} \right). \hspace{11mm}
\label{trans_dgvYO-3} 
\end{subnumcases}
In the more general case where 
\mbox{$\lim_{n \to - \infty}\left( \sca{\vt{S}_{n}}{\widebar{\vt{S}}_{n-1}} + \sca{\vt{S}_{n-1}}{\widebar{\vt{S}}_{n}} \right) \neq0$}, 
we 
need 
only 
replace 
$\prod_{j=-\infty}^n v_j$ 
in (\ref{v_prod}) and (\ref{trans_S_r}) 
with 
\[
\left( \prod_{j=-\infty}^n v_j \right) 
\exp \left[ \frac{1}{2} \int \lim_{n \to - \infty}
 \left( \sca{\vt{S}_{n}}{\widebar{\vt{S}}_{n-1}} + \sca{\vt{S}_{n-1}}{\widebar{\vt{S}}_{n}} \right) \mathrm{d} t \right]. 
\]
Note, incidentally, that  (\ref{trans_dgvYO}) can be rewritten in a slightly simpler form 
if we use the new variable \mbox{$u_n := v_n v_{n+1}$} instead of $v_n$. 

The transformed system (\ref{trans_dgvYO}) admits two interesting reductions. 
First, 
by setting \mbox{$v_n=0$}, 
(\ref{trans_dgvYO}) reduces to 
\begin{equation} 
\nonumber
\left\{ 
\begin{split}
& \mathrm{i} \vt{r}_{n,t} = \vt{r}_{n-1} 
	-\frac{\mathrm{i}}{2} \vt{r}_n \sca{\vt{r}_{n}}{\widebar{\vt{r}}_{n+1}} -c \vt{r}_{n}, 
\\[1pt]
& \mathrm{i} \widebar{\vt{r}}_{n,t} = -\widebar{\vt{r}}_{n+1} 
 +\frac{\mathrm{i}}{2} \widebar{\vt{r}}_n \sca{\vt{r}_{n-1}}{\widebar{\vt{r}}_{n}} + c \widebar{\vt{r}}_{n}, 
\end{split}
\right.
\end{equation}
which is essentially an elementary flow of a 
discrete integrable 
nonlinear Schr\"odinger hierarchy~\cite{GI2}; 
this discrete 
nonlinear Schr\"odinger hierarchy is closely related to an elementary auto-B\"acklund
transformation for the continuous nonlinear Schr\"odinger hierarchy 
as clarified by 
Konopelchenko~\cite{Kono82} and 
D.~V.\ and G.~V.~Chudnovsky~\cite{Chud1,Chud2}.  
Second, by setting 
\begin{equation}
\vt{r}_n = \vt{r}, \hspace{5mm} \widebar{\vt{r}}_n = \widebar{\vt{r}}, \hspace{5mm}  
\sca{\vt{r}}{\widebar{\vt{r}}} = -2 \mathrm{i}, \hspace{5mm} c=0, 
\nonumber 
\end{equation}
(\ref{trans_dgvYO}) reduces to 
the modified version of the Volterra lattice~\cite{Wadati76}: 
\begin{equation}
v_{n,t} = -\mathrm{i} v_n^2 \left( v_{n+1} - v_{n-1} \right), 
\nonumber
\end{equation}
which coincides with (\ref{mLV}) up to a rescaling of 
the 
variables. 

\section{Concluding remarks}

In this paper, we have shown that the discrete multicomponent Yajima--Oikawa system (\ref{dgvYO}) 
admits the Lax-pair representation, which then can be used to generate an infinite number of conservation laws 
in a recursive manner. Thus, the integrability of the discrete multicomponent Yajima--Oikawa system (\ref{dgvYO}) 
is established. 
The next higher flow 
of the discrete multicomponent Yajima--Oikawa 
hierarchy is presented, 
which can be reduced to the modified version of the Volterra lattice. 
The discrete multicomponent Yajima--Oikawa system (\ref{dgvYO}) 
turns out to be related 
to (the vector generalizations of) two integrable discrete 
nonlinear Schr\"odinger hierarchies. 
First, 
in 
a 
suitable scaling limit, 
(\ref{dgvYO}) can 
be reduced to 
an 
elementary 
flow of 
the Ablowitz--Ladik hierarchy. 
Second, by applying a nonlocal transformation 
and imposing a reduction, 
(\ref{dgvYO}) 
can be simplified 
to 
an elementary flow of the 
Konopelchenko--Chudnovsky hierarchy 
as well as the modified version of the Volterra lattice. 

In a suitable continuous limit, 
the Lax-pair 
representation 
for the discrete multicomponent Yajima--Oikawa system 
(\ref{dgvYO})  
can be reduced to the 
Lax-pair representation for the continuous multicomponent Yajima--Oikawa system. 
However, 
the 
size 
of the Lax 
matrices in the discrete case is 
larger 
than that in the continuous case
(cf.~Proposition~\ref{prop2.1} and 
Proposition~\ref{prop2.4}); 
thus, it is not possible to find the Lax-pair 
representation 
in the discrete case 
directly 
by discretizing the Lax-pair representation 
in the continuous case. 

%
%

\end{document}